\documentclass[authoryear,final,3p,times,twocolumn]{elsarticle}

\usepackage{amssymb}

\usepackage{url}

\usepackage{geometry}
\usepackage{fancyhdr}
\journal{JQSRT}

\begin{document}

\begin{frontmatter}



\title{Theory and practice of simulation of optical tweezers}

\author{Ann A. M. Bui}
\author{Alexander B. Stilgoe}
\author{Isaac C. D. Lenton}
\author{Lachlan J. Gibson}
\author{Anatolii V. Kashchuk}
\author{Shu Zhang}
\author{Halina Rubinsztein-Dunlop}
\author{Timo A. Nieminen}
\ead{timo@physics.uq.edu.au}

\address{The University of Queensland,
School of Mathematics and Physics,
Brisbane QLD 4072, Australia}

\address{\large Journal of Quantitative Spectroscopy and Radiative Transfer
\textbf{195}, 66-75 (2017)}

\begin{abstract}
Computational modelling has made many useful contributions to
the field of optical tweezers. One aspect in which it can be applied
is the simulation of the dynamics of particles in optical tweezers.
This can be useful for systems with many degrees of freedom, and for
the simulation of experiments. While modelling of the optical force
is a prerequisite for simulation of the motion of particles in
optical traps, non-optical forces must also be included;
the most important are usually Brownian motion and viscous drag.
We discuss some applications and examples of such simulations.
We review the theory and practical principles of simulation of
optical tweezers, including the choice of method of calculation
of optical force, numerical solution of the equations of motion
of the particle, and finish with a discussion of a range of
open problems.
\end{abstract}


\end{frontmatter}


\section{Introduction}

The optical forces in optical tweezers result from the interaction
of the trapping beam with the trapped particle. Thus, the computation
of optical forces and torques is a light scattering problem. While this
is a challenging problem, and much work remains to be done, there has been
a great deal of progress, and for  many situations, it is straightforward to
obtain the optical force and torque. However, if we wish to simulate the
behaviour of particles within optical tweezers, the optical force is only
one of the necessary ingredients. We will discuss some applications
and examples of such simulations, and review the theory and principles
of simulation of optical tweezers.

\subsection{The need for simulations}

Since it is usually straightforward to calculate the optical force
on a trapped particle, it is possible to characterise the trap by
determining the force as a function of particle position
(and orientation if the particle is non-spherical). At first glance,
this appears to provide complete information about the trap, and
we might ask what need there is to perform simulations. There are
two main answers to this question. First, it is not always feasible
to generate such a force map of the trap. Second, while a force map
of this type does contain complete information about the trap in some
sense, it doesn't directly answer all questions we might have
about the trap. In particular, the dynamics of a particle in the
trap depend on its interaction with the surrounding environment
as well as the optical force. The dominant elements of that interaction
are often Brownian motion and viscous drag, but other types
of interaction can also be important. Where the dynamics themselves
are the object of study (e.g., escape probabilities, synchronised
dynamics of trapped particles, etc.) or have a major impact on
the behaviour of interest (e.g., in the simulation of
measurements to test calibration procedures), it is necessary
to take these non-optical into account.

The first of these cases results from situations with many degrees
of freedom. To map the force as a function of position with
useful (but not high) resolution typically requires about 30 steps
along each degree of freedom (giving about 10 steps as forces change from
zero to a maximum value). If it takes 1 second to calculate the optical
force at a single position, this will give required computational
times for different degrees of freedom (DOF) of:
\begin{description}

\item [1 DOF] Example: calculating axial and/or radial force--position
curves; finding equilibrium position along beam axis, and axial and
radial spring constants. 30 to 60 points. Time: 0.5--1 minute.

\item [2 DOF] Example: mapping force for a spherical particle in a
rotationally symmetric trap (e.g., circularly polarised Gaussian beam).
$30^2 \approx 1000$ points. Time: $\approx$ 15 minutes.

\item [3 DOF] Example: mapping force for a spherical particle in a
trap lacking rotational symmetry (e.g., linearly polarised Gaussian beam).
$30^3 \approx 30,000$ points. Time: $\approx$ 8 hours.

\item [4 DOF] Example: mapping force for a rotationally symmetric
non-spherical particle in a
rotationally symmetric trap.
$30^4 \approx 10^6$ points. Time: $\approx$ 10 days.

\item [5 DOF] Example: mapping force for a rotationally symmetric
non-spherical particle in a
trap lacking rotational symmetry; two spherical particles in a
rotationally symmetry trap.
$30^5 \approx 3\times 10^7$ points. Time: $\approx$ 1 year.

\item [6 DOF] Example: mapping force for a non-spherical particle
lacking rotational symmetry in a
trap lacking rotational symmetry; two spherical particles in a trap
lacking rotational symmetry.
$30^6 \approx 10^9$ points. Time: $\approx$ 30 years.

\end{description}

Additional particles will add 2--3 translational degrees of freedom
(depending on the symmetry of the trap) and 0, 2, or 3 rotational
degrees of freedom (depending on the symmetry of the particle).
If the trapping beam varies in time, this adds another degree of freedom,
although if the time variation consists of switching between a small
number of fixed positions, this will only multiply the number of required
calculations and the computational time by a small number.

The above times do not take parallelisation of the calculations
into account---this can readily bring one or two more degrees of
freedom into feasibility. However, even with parallelisation, we
will still rapidly run into the limits of practicality due to the
exponential growth of computational time with the number of
degrees of freedom. Therefore, it can be necessary to resort to
simulation to obtain information we might prefer to find from
a complete force map. This will typically involve non-spherical
particles or multiple particles.

On the other hand, even if it is feasible to calculate a complete
force map for the trap, we might still wish to perform simulations.
In particular, a force map doesn't contain information about the
dynamics of a particle in the trap. While the optical force---which
the force map provides---is a key factor in the dynamics of
the particle, the particle is also influenced by other forces:
viscous drag, thermal forces (driving Brownian motion),
and possibly interaction with other parts of the
environment. If the dynamics are of interest, we can use simulation
to uncover it.

To explore the dynamics of a particle in the trap, it can be possible,
and advantageous, to use a pre-calculated force map.
If it is feasible to calculate a complete force map with reasonable
resolution, the optical force at any position can be found by
interpolating between the points in the force map where the forces
are known. This interpolation can be performed very quickly
(the computational implementation should avoid copying the force map
to perform the interpolation). The required accuracy of the interpolated
force will determine the minimum resolution of the force map. This resolution
of the force map, along with the
required spatial extent of the simulation, determines the number of points
required in the force map. If this exceed the number of time steps required
in the simulation, then direct calculation will be more efficient. However,
often the number of time steps will be much greater, and using a force
map to find the optical force will be much more efficient. This will often
be the case for optical traps with 2 or 3 degrees of freedom. An extreme case
of this is where the particle remains very close to its equilibrium position,
and the trap can be represented in terms of a spring constant (which will
generally be a diagonal tensor, with different spring constants in different
directions, or even a non-diagonal tensor).

\subsection{Applications of simulations}

There are many possible applications of this. Most fall into three broad
categories: simulations to understand experiments that have been
performed, simulations to predict the results of potential experiments,
and simulations to explore optical traps and the dynamics of trapped
particles in ways that are not accessible experimentally.

The first of these, simulations of experiments that have been
performed, can be simply seeing if a simulated experiment matches
measured results. This can be very useful if the experimental
results are surprising. If agreement between simulated and measured results
is obtained, the physics and models used in the simulation adequately
model reality. If agreement is not obtained, then the model is
either incomplete (e.g., physics not included significantly affect the
measured results) or elements of the model are incorrect (invalid approximations,
mathematical errors, incorrect implementation in software, numerical errors).

For example, \cite{stilgoephase} observed the appearance of a third trapping
equilibrium position as two optical traps were moved close together. In
this case, simulations were valuable for confirming that the third trap can
be produced in this two-beam configuration, even if the two trapping
beams are not mutually coherent, i.e., the third trap doesn't depend
on interference between the two trapping beams.

\cite{haghshenasjaryani2014} used a combination of experiment and
simulation to explore the transition from overdamped motion to
underdamped motion as the size of trapped particles was reduced.

\cite{volpespeckle} observed the trajectories of particles in a laser
speckle pattern, and compared the observed trajectories with simulated
trajectories in speckle patterns with the same average intensity.
In this case, the simulations do not aim to exactly replicate
the experimental situation, but to replicate it in a statistical
sense. That is, the speckle fields in the simulations have the same
statistical properties as the experimental speckle fields. Qualitative
and statistical agreement between observed and simulated trajectories
demonstrates that the observed behaviour is general, and does not
arise due to some abnormality in the experimental case.

\cite{wunonconservative} determined the non-conservative force field
from the motion of a trapped particles. Their experimental results
were supported by simulations. In the experiment, the force field is
inferred from the motion of the particle, while in the simulations,
the force field is known. This allows validation of the procedure used
to obtain the experimental force field.

Similarly, the knowledge of the optical forces available in simulations
was used to validate escape force calibrations on chromosomes
by \cite{nima_sr}. Measurements of the force required to pull
chromosomes free from an optical trap were performed in order to
estimate the forces exerted on chromosomes by a cell during cell division
(mitosis) from the power required to halt the motion \citep{ferrarogideon}.
Since the exact size and refractive index of the chromosomes were uncertain,
a further series of experiments and simulations on the escape of
spheres from optical traps due an applied force were performed \citep{ann_spheres},
revealing the dependence of the escape trajectory, and the escape force, on
the trapping power and rate of increase of the applied force.

Simulations that deliberately differ from the experiments can
show the effect of the difference on observations or measurements.
For example, \cite{czerwinski} used Allan variance to quantify
noise in optical tweezers setups. Simulations were used to obtain
(simulated) data free of noise and long-term drift, providing a
suitable baseline for comparison with experimental results.

As noted above, simulations are often necessary when the system
has many degrees of freedom, such as when there are non-spherical
or multiple particles. \cite{brzobohaty} used simulations to
explore the shape dependence of the trapping behaviour of
non-spherical gold nanoparticles. \cite{brzobohaty2} used
simulations to support experiments rotational dynamics of
multiple spheroidal particles in a dual beam trap. 
Following observations of optically-driven oscillations
of ellipsoidal particles \citep{mihiretie},
\cite{loudet} performed simulations to understand the
physical basis of the observed behaviour.

In these examples above, simulations were performed to support
experiments. The opposite of this, where experiments are performed
to support simulations, is also common. The aim of the simulations
can vary greatly, from demonstration of the feasibility of a
particular experiment before performing it, to using simulation as a
tool to help design the experiment, through to a broad series
of exploration via simulation with experiments being performed
to validate the simulations. In this last case, the experimental
work might consist of only a small fraction of the range covered
by the simulations. If the simulation method and implementation is already
known to be reliable from previous validation, then the reported work
might consist purely of simulations.

Some of this more simulation-focussed work is similar to the experiment-focussed
work described above. For example, 
similar to the work of \cite{wunonconservative},
\cite{pesce_epl} also explored the non-conservative forces in optical traps.

As noted above, simulations allow the optical forces to be known,
and are therefore valuable for testing calibration methods. Simulated
measurements, of the type that would be measured experimentally in
order to calibrate an optical trap, or to determine the optical force
field from the motion of a trapped particle, can be generated, and
the same analysis that would be performed on experimental data can
be performed on the simulated data. The simulated calibration or
force measurement can be compared with the actual optical force in
the simulation. Examples of simulations of this type include
\cite{volpe_pfm}, \cite{volpe_torque}, and \cite{gong_moc}. Such
simulations can also readily include non-spherical particles
\citep{bui}. Similar comparison on known quantities in the simulation
and simulated measurement of these quantities can be carried out
for methods to measure properties of the surrounding medium,
such as its viscoelasticity \citep{fischer2007}.

An application where the trapping beam varies in time is the
simulation of control methods, where the position or power of the
beam can be varied to achieve a desired effect
\citep{banerjee_control,aguilaribanez,li_control}.
Variation of the beam power over time introduces an
additional degree of freedom, and movement of the beam introduces
2 to 4 additional degrees of freedom (time and 1 to 3 spatial degrees
of freedom). Similarly, improved trapping methods can be explored
\citep{taylor}.

Simulations can be aimed at a more general exploration of the
behaviour of optically-trapped particles. These can be specifically
investigating the dynamics of trapped particles
\citep{banerjee_sim,xu_sim,deng_sim,ren10,cao2016,trojek}.
Another common goal is the study of the behaviour of
non-spherical particles, where the additional degrees of freedom
motivate the use of simulations
\citep{simpson10,simpson11,cao2012}. These can include optically-driven
micromachines. One example is the use of simulations to determine
the optimum illumination to drive a corrugated rotor with maximum
torque efficiency, while retaining stable three-dimensional trapping
\citep{loke_donut}. Another example is an optical ``wing'', 
consisting of a semi-cylindrical rod \citep{wing1,wing2},
Such a structure can generate lift---an optical force acting normal to the
direction of illumination---in addition to the expected radiation pressure
force. Simulations by \cite{wing3} show that complex rocking motion can occur.
The simulations allow the effects of time-varying illumination,
producing a parametrically driven nonlinear bistable oscillator, to be explored.

Finally, simulations can be useful for educational purposes
\citep{volpe_ajp,perkins}

For some of these simulations, it is not necessary to accurately
model the dynamics of the particle. For example, to determine the equilibrium
position and orientation of a non-spherical particle within the trap,
it is not necessary to correctly model the viscous drag.
The translational and
rotational drag tensors can be approximated by Stokes drag for a sphere, even
if the particle is non-spherical. Brownian motion can be ignored, although
it (or random jitter providing similar random motion) can be useful
for preventing the particle from getting stuck
in an unstable equilibrium. This can happen, for example, if the particle is
a flat disc, which would tend to align with its symmetry axis normal to the beam
axis \citep{bayoudh2003}, but will be in an unstable equilibrium if the simulation is begun with
the disc on the beam axis, with its symmetry axis along the beam axis.

However, for many types of simulations, where actual or prospective experiments are being
simulated, it is often important to accurately model the dynamics, and
to include all important details of the interaction of the particle with
its environment. At minimum, this can be expected to include viscous drag
and Brownian motion.

\section{A recipe for simulation, part 1}

The motion of a particle of mass $m$ subject to a force $\mathbf{F}(t)$ can
be calculated from
\begin{equation}
m\frac{\mathrm{d}^2\mathbf{r}}{\mathrm{d}t^2} = \mathbf{F}(t),
\label{ode_newton}
\end{equation}
given initial conditions for the position $\mathbf{r}(t=0)$ and velocity
$\mathbf{v}(t=0)$. The force $\mathbf{F}(t)$ is the sum of contributions
from various sources:
\begin{eqnarray}
\mathbf{F}(t) & = & \mathbf{F}_\mathrm{optical} + \mathbf{F}_\mathrm{weight} +
\mathbf{F}_\mathrm{buoyancy} \nonumber \\
 & & + \mathbf{F}_\mathrm{drag} +
\mathbf{F}_\mathrm{Brownian} + \mathbf{F}_\mathrm{other},
\end{eqnarray}
where we have explicitly listed the optical force, weight, buoyancy, viscous
drag, and thermal forces driving Brownian motion. We have also included
$\mathbf{F}_\mathrm{other}$ to represent any other forces present. The
weight and buoyancy are straightforward, with
\begin{equation}
\mathbf{F}_\mathrm{weight} = mg = \rho_\mathrm{particle} V\mathbf{g}
\end{equation}
and
\begin{equation}
\mathbf{F}_\mathrm{buoyancy} = \rho_\mathrm{medium} V \mathbf{g},
\end{equation}
where $V$ is the volume of the particle, $\rho_\mathrm{particle}$
and $\rho_\mathrm{medium}$ are the densities of the particle and
the surrounding medium, and $\mathbf{g}$ is the local gravitational
acceleration. These can almost always be treated as constant,
and present no difficulty for numerical solution of the differential
equation~(\ref{ode_newton}).

The optical force $\mathbf{F}_\mathrm{optical}$ is the force
that typically has the most attention paid to it in discussion
of optical tweezers and optical trapping. Depending on the particle
in question (and the optical trap), calculation of the optical
force can vary from a formidable computational challenge to
an already-solved problem with freely-available implementations.
We will discuss the calculation of optical forces in the following
section. For the moment, we will consider the spatial scale over
which the optical force varies, since this directly affects the
solutions of differential equation~(\ref{ode_newton}). In a typical
optical trap, the energy density varies from small to large
values over a distance of half a wavelength or more. As a result,
the optical forces will vary from small to large over a similar
length scale, or over a distance comparable to the particle
radius. For example, when a large spherical particle is centred
on the beam axis in typical Gaussian beam optical tweezers, the force
is zero, and when the edge of the particle is on the beam axis, the
radial force is approximately maximum. For such a trap, we can
assume that the length scale over which the optical force
varies is the larger of the two (i.e., the maximum of the
half-wavelength and the particle radius). However, even for large
particles, the force can vary over the half-wavelength scale,
if interference effects are important, such as when trapping
in interference fringes produced by mutually-coherent
counter-propagating beams. Knowledge of the length scale of
variation in the optical force allows us to estimate a
suitable maximum distance to allow the particle to move
over a time step when numerically solving
equation~(\ref{ode_newton}).

The viscous drag force has major effects on the numerical
solution of the differential equation describing the motion.
We will discuss details of the calculation of the viscous
drag later, and restrict the current discussion to 
these effects on the solution. Since, typically, trapped
particles are microscopic and are trapped in a viscous
environment, the interaction between the particle and the
fluid is characterised by very low Reynolds numbers.
In this case, the viscous drag is linearly related to
the velocity:
\begin{equation}
\mathbf{F}_\mathrm{drag} = \mathbf{D} \mathbf{v},
\label{drag}
\end{equation}
where $\mathbf{D}$ is a $3\times 3$ drag tensor.
Commonly, it is simply stated that since the Reynolds
number is very low, we can neglect the inertial
term in equation~(\ref{ode_newton}) (recalling that the
Reynolds number is the ratio of inertial effects to viscous
effects in fluid motion), reducing equation~(\ref{ode_newton})
to
\begin{equation}
0 = \mathbf{F}(t).
\label{ode_aristotle1}
\end{equation}
Substituting (\ref{drag}), we obtain the first-order differential
equation
\begin{eqnarray}
 - \mathbf{D} \frac{\mathrm{d}\mathbf{r}}{\mathrm{d}t} & = &
\mathbf{F}_\mathrm{optical} + \mathbf{F}_\mathrm{weight} +
\mathbf{F}_\mathrm{buoyancy} \nonumber \\
& & + \mathbf{F}_\mathrm{Brownian} + \mathbf{F}_\mathrm{other}.
\label{ode_aristotle2}
\end{eqnarray}
Note that differential equation~(\ref{ode_aristotle2}) assumes
that the particle is always moving at terminal velocity, with
viscous drag balancing the sum of the other forces. When is this
condition satisfied? If a particle is initially at rest in a fluid,
and a force $\mathbf{F}$ is suddenly turned on, the approach to terminal
velocity $\mathbf{v}_0 = -\mathbf{D}^{-1}\mathbf{F}$
will be characterised by a time constant $\tau$ such that
\begin{equation}
\mathbf{v}(t) = \mathbf{v}_0 (1-\exp(-t/\tau)).
\end{equation}
The time constant $\tau$ is
\begin{equation}
\tau = m |\mathbf{v}_0|/|\mathbf{F}|,
\end{equation}
which, for a spherical particle, becomes
\begin{equation}
\tau = 2 \rho_\mathrm{particle} a^2 / (9\eta),
\end{equation}
where $a$ is the radius of the particle, and $\eta$ is the (dynamic)
viscosity of the surrounding medium. Notably, this is independent of
the force. For a particle of radius $a = 1\,\mu$m in water, this
gives a time constant of $\tau \approx 2.4\times 10^{-7}$s.
If we are calculating the motion of the particle using a time step
$\Delta t$ large compared to this (e.g., $\Delta t > 10\tau$), we can
safely use equation~(\ref{ode_aristotle2}). If we are using
a time step similar to or smaller than $\tau$, we should,
strictly speaking, use equation~(\ref{ode_newton}). In practice,
if the force only changes by a small amount over the time step
$\Delta t$, the particle will already be moving at close to
terminal velocity, and equation~(\ref{ode_aristotle2}) will
yield acceptable results. \cite{haghshenasjaryani2014} consider
a case where the difference between equations (\ref{ode_newton})
and (\ref{ode_aristotle2}) matters over short times.

Motion at the very low Reynolds numbers typical in optical traps
is outside our everyday experience. \cite{purcell1977b} gives
an excellent and accessible overview.

Brownian motion presents a serious difficulty: numerical solution
of differential equations such as (\ref{ode_newton})
and (\ref{ode_aristotle2}) typically depend on using a time step
sufficiently short so that the time-varying quantities in the
equations (the forces, the position, and the velocity) only vary
by a small amount over the time step. However, no matter how short
a time step is chosen, classical Brownian motion (i.e., Brownian
motion in a continuous fluid \cite{einsteinbook}) always varies by
a large amount. Therefore, it is simplest to remove Brownian
motion from our differential equations, and treat it separately.
We will return to this point after discussion of optical forces,
viscous forces, and Brownian motion.

The final force in our differential equations,
$\mathbf{F}_\mathrm{other}$, can represent many possible forces.
For example, adhesion forces between particles or particles and
the microscope slide, electrostatic forces, magnetic forces,
and more. Needless to day, some of these forces can be
challenging to model accurately. In some cases, a similar approach
to that suggested above for Brownian motion will be useful: remove
the force from the differential equation, and treat it separately.

Finally, for non-spherical particles and some spherical particles,
it is necessary to consider rotational motion as well as translational
motion. In this case, our differential equation will include
optical torques, viscous drag torque (which will be linearly related
to the angular velocity by a rotational drag tensor), and other torques.
These can be treated in a similar manner to their translational
counterparts. Rotational Brownian motion can be treated in a similar
manner to translational Brownian motion. Weight and buoyancy can
often be ignored, since for many particles, they produce no torque about
the centre of the particle. One complication is that it is often
desirable to perform the calculations of optical force and torque
in a coordinate system fixed to the particle, necessitating transformations
between the particle frame and the stationary frame.
Due to the analogous nature of rotational motion compared with translational
motion, the rotational equations of motion can be readily written
following the translational equations of motion, replacing
masses with moments of inertia, forces with torques, and translational
drag tensors with rotational drag tensors.
It should be noted that for particles with a chiral shape, rotational and
translational motion can be coupled through viscous drag \cite{moffatt1977};
in this case, an addition coupling tensor will be included in both the
translational and rotational equations of motion.
For descriptions of simulations
involving rotational motion, including equations of motion, see \cite{cao2012,bui}.

\section{Optical forces and torques}

The computational modeling and simulation of optical tweezers is essentially a
light scattering problem \citep{nieminen2001jqsrt,nieminen2014}.
The trapping beam interacts with the trapped
particle---this is the scattering aspect of the problem---and a force results.
As there are a large number of computational approaches to light scattering
\citep{kahnert2003b},
there are a large number of computational approaches to calculating optical
forces  \citep{nieminen2014,jonesbook}. A complete review of all of the methods
would be a monumental (and book-length, if not multi-volume) task, and we will
not attempt it here. Instead, we will discuss the elements of calculation
of optical forces that are most important for deciding which method
will be used for such calculations, and refer readers to appropriate
technical literature for particular methods.

We will begin with an overview of the T-matrix method and
why it is often the method of first choice for simulations. Note that
for a spherical particle, the T-matrix method is essentially
equivalent to generalised Lorenz--Mie theory (GLMT) \citep{gouesbet2010oc}.
Then, after noting cases where the T-matrix method might not
be the best choice, or even feasible, we review some basic
principles of calculating optical forces that can affect
the choice of alternative methods.

In general, it is safe to conclude that where calculation of
the T-matrix is feasible, the T-matrix method appears to be
the ideal method. The T-matrix method is \emph{not} a method
of calculating light scattering by a particle, but a formalism
in which the \emph{already calculated} scattering properties
of the particle can be expressed in the form of the T-matrix.
The extended boundary condition method (EBCM) is widely used
to calculate the T-matrix, being the original method used
by \cite{waterman1965,waterman1971}. Thus, ``T-matrix method''
is often used synonymously with EBCM, but the distinction
between them should be recognised \citep{nieminen2014,gouesbet2010oc,gouesbet2015}.

In the T-matrix method, we represent the incident
and scattered fields in terms of discrete sets of
vector-valued basis functions
$\psi_n^{(\mathrm{inc})}$ and $\psi_n^{(\mathrm{scat})}$,
where $n$ is a mode index labelling the functions,
each $\psi_n$ being a solution of the vector Helmholtz equation.
Using these bases, we can write the incident field amplitude as
\begin{equation}
\mathbf{E}_0^{(\mathrm{inc})} = \sum_n^\infty a_n \psi_n^{(\mathrm{inc})},
\label{exp1}
\end{equation}
where $a_n$ are the \emph{mode amplitudes}
(or \emph{expansion coefficients}) of the incident wave, and
the scattered wave amplitude as
\begin{equation}
\mathbf{E}_0^{(\mathrm{scat})} = \sum_k^\infty p_k \psi_k^{(\mathrm{scat})},
\label{exp2}
\end{equation}
where $p_k$ are the mode amplitudes of the scattered wave.
For computational practicality, these sums must be truncated
at some finite $n_\mathrm{max}$. For a basis set of vector spherical
wavefunctions, as usually used in the T-matrix method, the
truncation criterion given by Brock is suitable, giving
a relative error due to truncation of about $10^{-6}$ \citep{nieminenjmo}.

With truncation, the mode amplitudes of the incident and scattered waves
can be written as column vectors $\mathbf{a}$ and $\mathbf{p}$, and
their relationship can be expressed in matrix form as
\begin{equation}
\mathbf{p} = \mathbf{T} \mathbf{a},
\label{t-matrix}
\end{equation}
where $\mathbf{T}$ is the \emph{T-matrix}, or transition matrix, or
system transfer matrix. This assumes that the electromagnetic properties
of the particle are linear and constant (i.e., the particle doesn't
change over time). With this description of scattering, the scattering
properties of the particle and the details of the incident field
are separated, leading to high efficiency for repeated calculation.
Once the T-matrix $\mathbf{T}$ has been calculated, it can be used
repeatedly for calculation under different illumination conditions
(as long as the wavelength remains the same). Thus, as the particle
moves within the optical trap, the T-matrix $\mathbf{T}$ remains
constant, and the incident field, described by $\mathbf{a}$ changes.
The changes in $\mathbf{a}$ can be found by using the transformation
properties of the basis functions $\psi_n^{(\mathrm{inc})}$ under
translation and rotation. It is important to note that the requirement
that the particle not change over time means that it is necessary to
use a coordinate system in which the particle is fixed. Thus, calculations
of the optical force and the torque are performed in the particle rest frame.

Further efficiency results from analytical integration of the
momentum flux, using known results for products of integrals of
the basis functions over a sphere. This reduces the formulae for
optical forces and torques from integrals to sums of products of
the mode amplitudes \citep{nieminen2014}. This avoids the need to calculate the
fields over a grid of points in order to perform such integral
numerically, and also avoiding numerical error due to the resolution
of the computational grid.

However, it is not always feasible to calculate the T-matrix of the
particle. The most common methods for calculating the T-matrix of
a particle are generalised Lorenz--Mie theory (GLMT), when the
particle is a uniform isotropic sphere (but note that GLMTs exist for
non-spherical particles as well \citep{gouesbet2015}) and 
the extended boundary condition method (EBCM), also known as 
the null-field method, developed by \cite{waterman1965,waterman1971}.
However, other methods are possible
\citep{mackowski2002,kahnert2003,nieminen2003b,gouesbet2010oc,mishchenko2010rev,loke2007a,loke}.
A comparison of EBCM, point-matching \citep{nieminen2003b}, and the discrete dipole
approximation (DDA) \citep{loke} is given by \cite{qi}.

Where it is impractical or impossible to calculate the T-matrix, other
methods can be sought.
In many ways, the calculation of optical forces and torques is a simple
scattering problem. Often, there is a single particle, comparable in size
to the wavelength, and sufficiently far from other particles and surfaces
so that multiple scattering can be ignored (it should be noted
that ``multiple scattering'' is a rather artificial concept
\citep{mishchenko2014book}, and it is
possible to treat scattering by a single object in a multiple scattering
formalism (e.g., using DDA) or scattering by a group of objects in a single scattering
formalism \citep{gouesbet1999h}). The incident field is
monochromatic and coherent. The complication is that the incident field
is not a plane wave, but a focussed beam. This, and the desired outputs
being the force and torque rather than the fields, or scattering cross-sections,
or scattering patterns, means that existing computational implementations of
particular methods might be unsuitable.

\begin{figure}[!htbp]
\centerline{\includegraphics[width=1.5in]{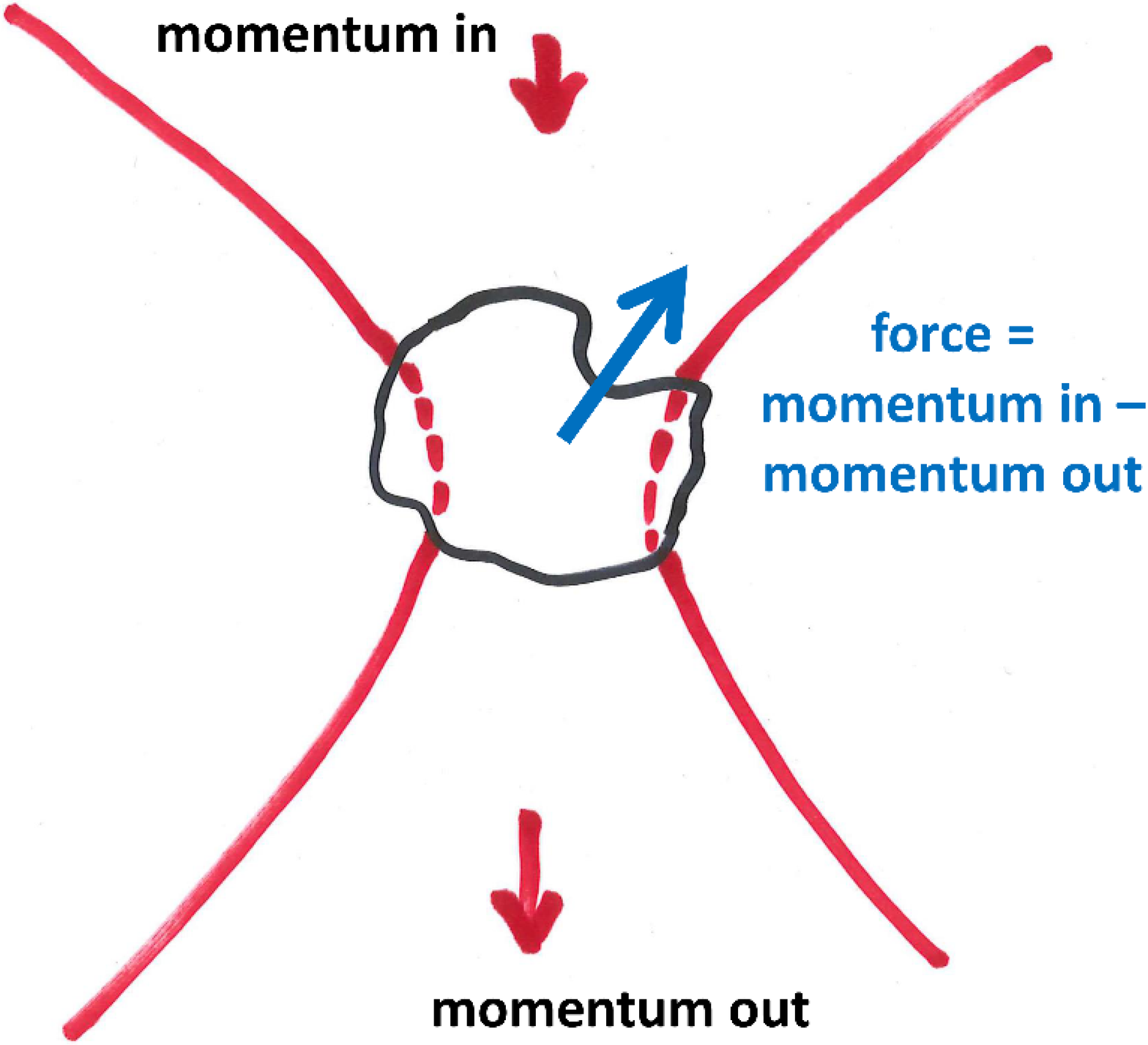}
\includegraphics[width=1.5in]{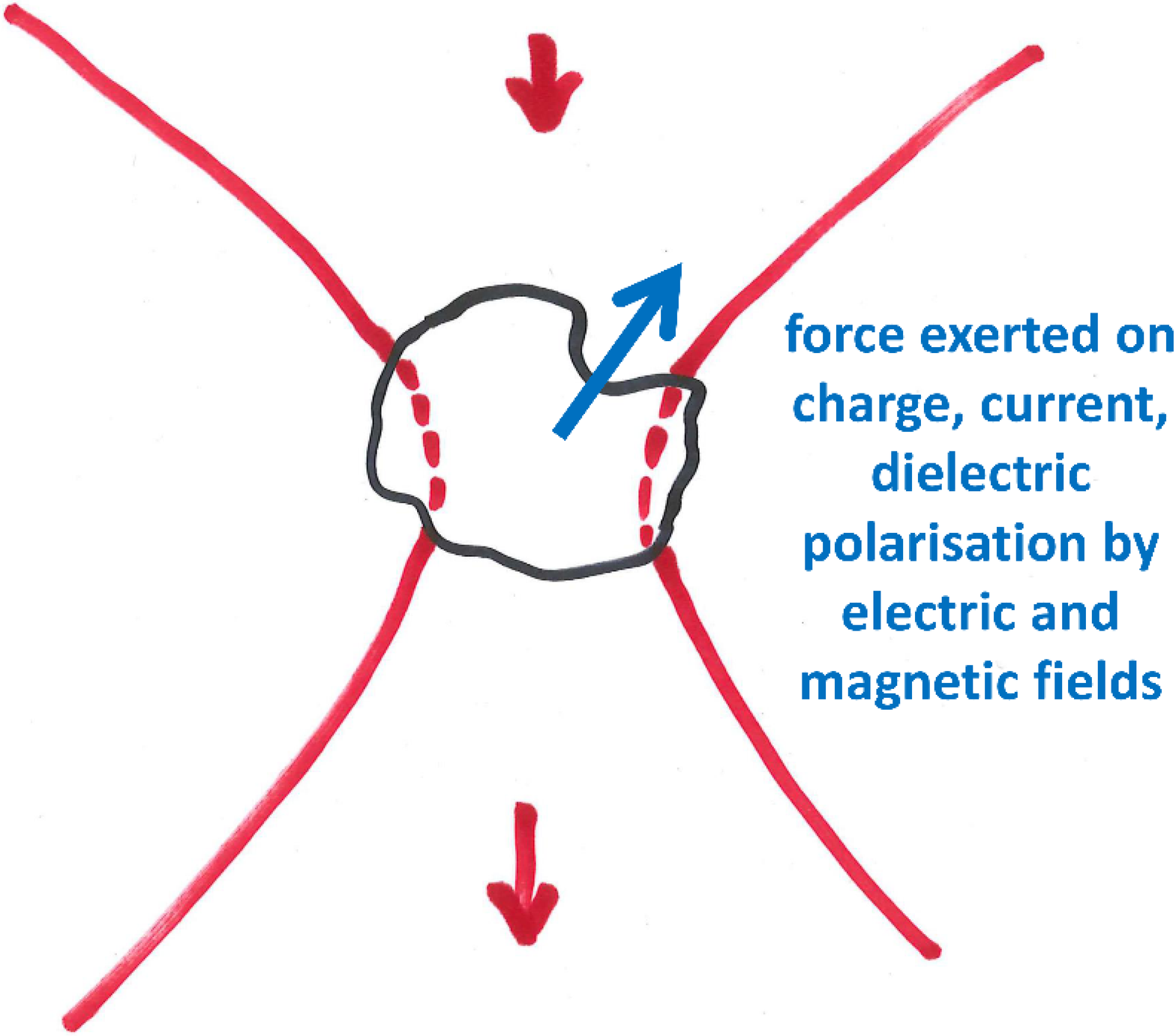}}
\caption{The optical force can be calculated
from the momentum flux or by an electromagnetic force
law.}
\label{flux_force}
\end{figure}

There are some general theoretical points that merit discussion, since they
can affect the choice of computational method or details of how a method is
implemented. In order to calculate the forces, there are two different
approaches that we can take. First, we can use conservation of momentum,
and find the difference between the incoming momentum flux and the outgoing
momentum flux of the light. This difference is the rate at which momentum
is transferred to the particle---that is, the force exerted on the
particle. Second, we can directly calculate the force using the
Lorentz force law (or the Helmholtz force law, or other force law).
These two approaches are summarized in figure~\ref{flux_force}.

At this point, one might be surprised to discover that there are
multiple expressions given in the literature for the momentum flux
of light, and also multiple electromagnetic force laws.
This is the Abraham-Minkowski controversy, where we encounter
competing expressions for the momentum density of an
electromagnetic field \citep{pfeifer2007}. With more than one possible expression
for the momentum density, how can we choose the correct one, or at least
the best one to use?

The key is to note that while there are different expressions for
the \emph{electromagnetic} momentum density in material media, the
different versions all correspond to identical expressions for the
\emph{total} momentum density in the material medium associated with
the electromagnetic wave. Where the electromagnetic momenta differ,
the difference is matched by opposing differences in what is labelled
material momentum or interaction momentum. Since we must calculate the
total force, whether or not it is described as purely electromagnetic
or the sum of an electromagnetic and a material force, the difference
in how the total momentum is divided into electromagnetic and material
(and possibly other) components is not fundamental. However, it is
convenient to be able to calculate a single quantity rather than
multiple quantities that, when added, equal that single quantity.
Noting that for cases where the electromagnetic properties of the medium
can be described completely with a constant permittivity and permeability,
the Minkowski momentum is the total momentum \citep{pfeifer2007}, this is the
simplest choice of momentum density.

\begin{figure}[!htbp]
\centerline{\includegraphics[width=1in]{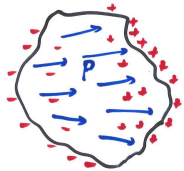}}
\caption{The optical force can be calculated
from the momentum flux or by an electromagnetic force
law.}
\label{force_laws}
\end{figure}

With each possible choice of momentum density, there is an associated
electromagnetic force law. If we begin by choosing a force law, we can
derive an expression for the momentum density and momentum flux of the
electromagnetic field. Doing this in reverse, we can begin with an expression
for the momentum, and obtain an electromagnetic force law.
The most commonly encountered force laws are the Lorentz force law, giving
the force acting on charges and currents, and the Helmholtz force law
which includes forces acting on induced dipole moments. These connection between
these force laws can be seen if we consider the induced dielectric
polarisation in a particle. We can represent this either by the dipole
moment per unit volume, or by equivalent charges. For a uniformly polarized
sphere, the dipole moment per unit volume is uniform throughout the sphere,
but we can replace this by an equivalent surface charge density. In the more
general case, we obtain a volume charge density from the variation of the dielectric
polarisation in the particle, as shown in figure~\ref{force_laws}.
The Helmholtz force law gives the force acting on
the dipole moment per unit volume, and the Lorentz force law the force acting
on the equivalent charges. In both cases, the total force is identical.

\begin{figure}[!htbp]
\centerline{\includegraphics[width=1.5in]{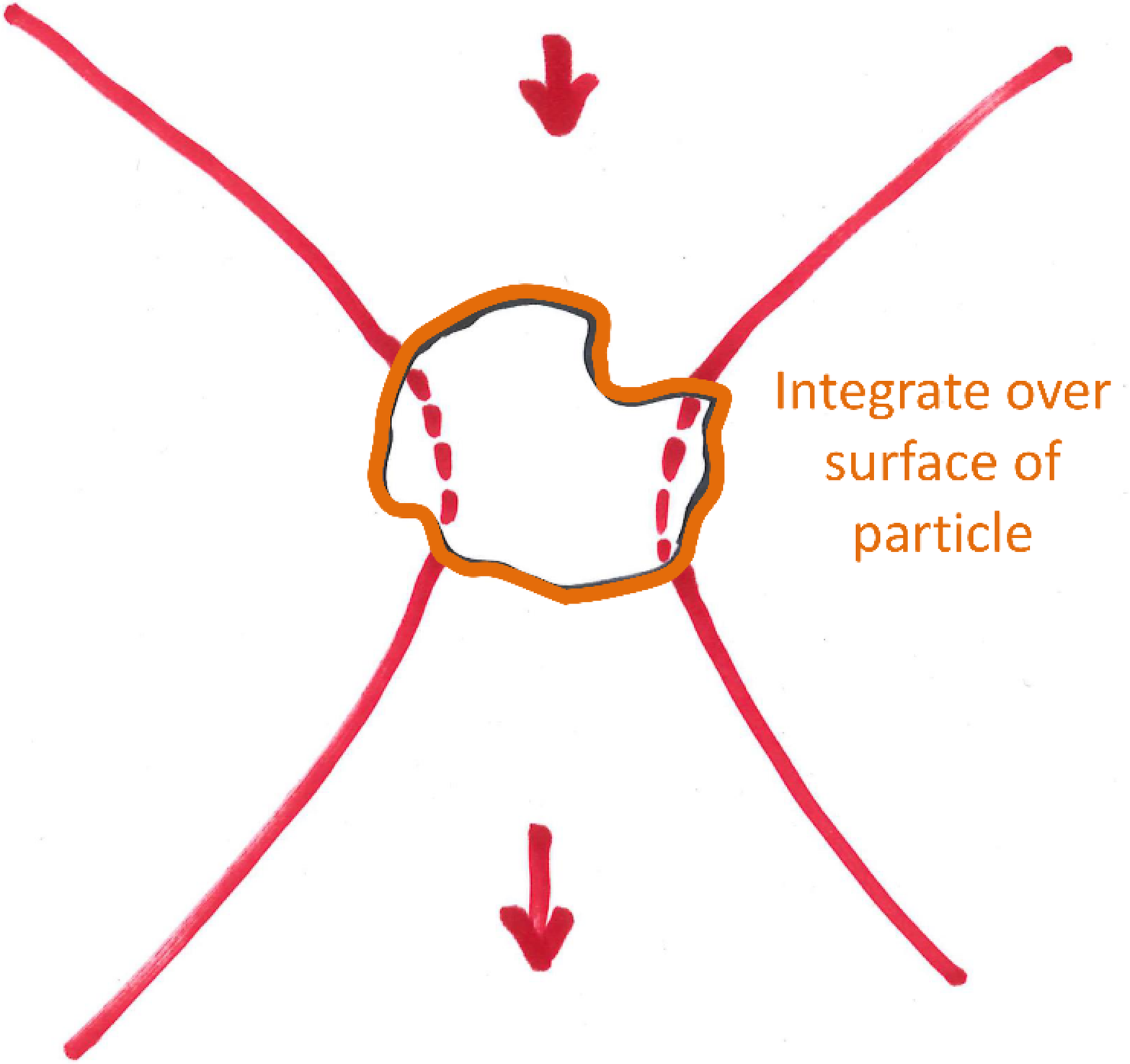}
\includegraphics[width=1.5in]{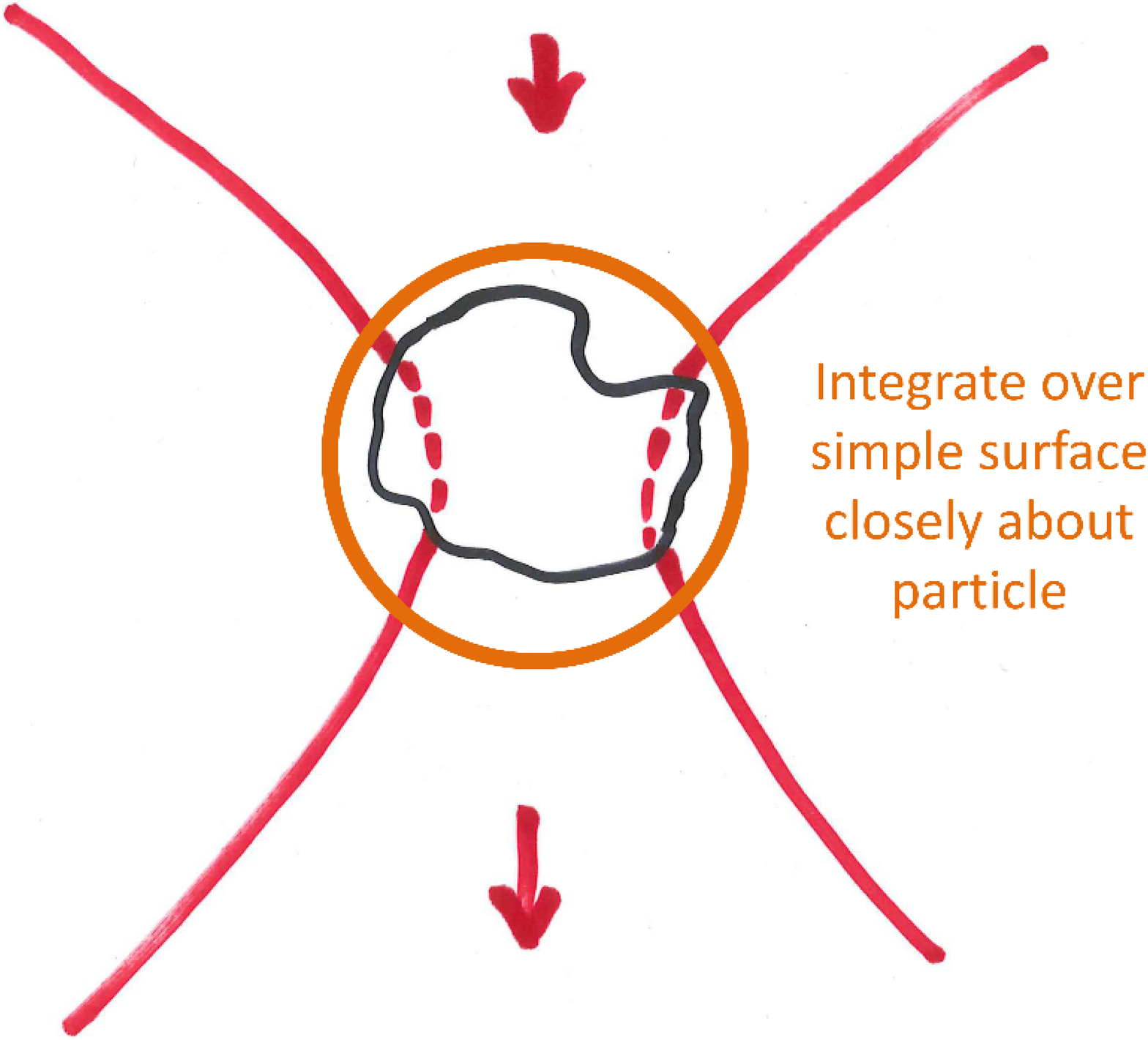}}
\centerline{\includegraphics[width=1.5in]{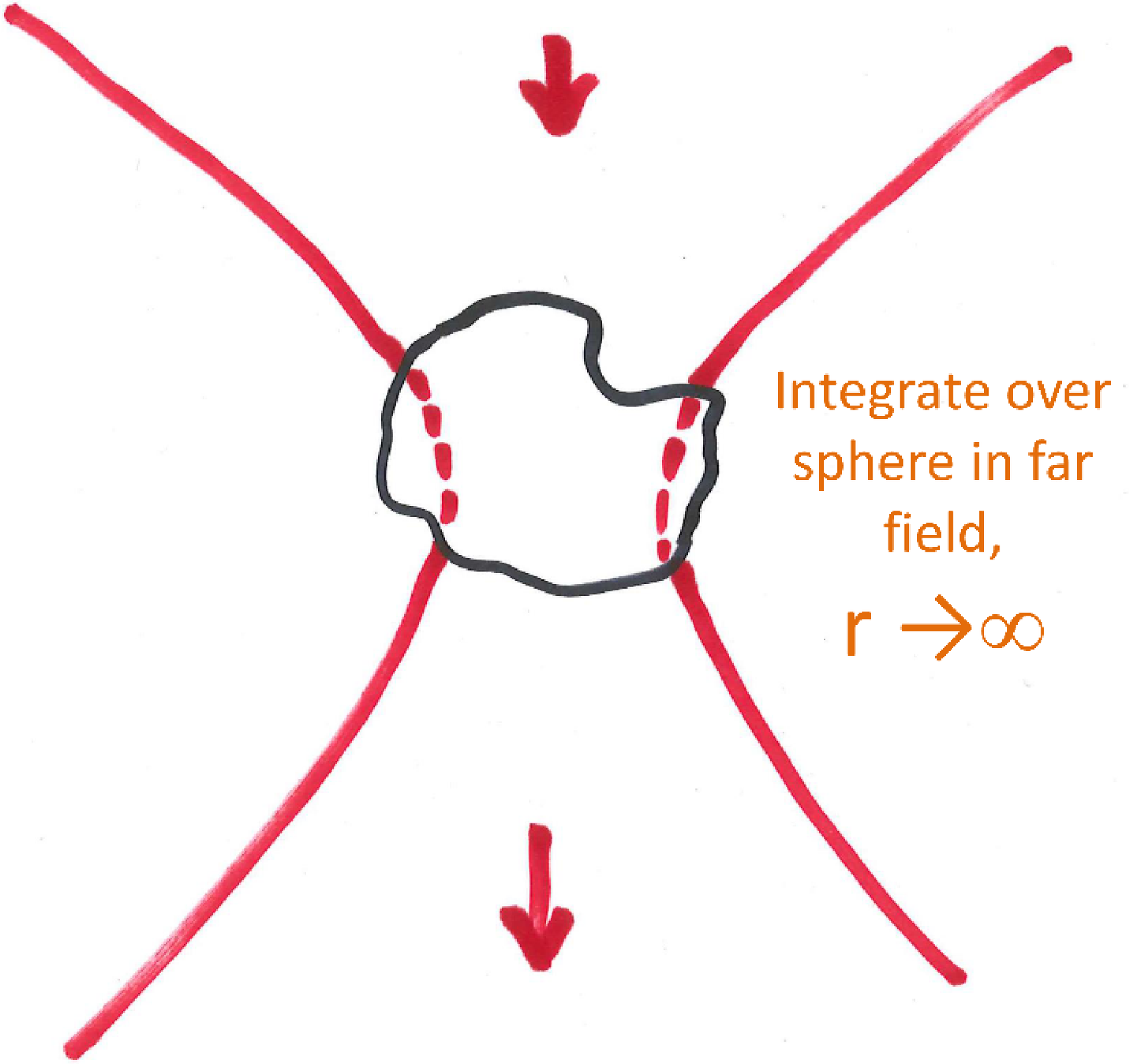}}
\caption{Choices of surface over which to integrate
momentum flux. We can choose a surface conforming to the surface of
the particle, or a surface of simple geometry enclosing the particle.
Alternatively, we can choose a spherical surface in the far field,
which can simplify the calculation by allowing us to make far-field
approximations.}
\label{integration_surface}
\end{figure}

If we choose to find the force from conservation of momentum, we need to choose
a closed surface over which to integrate the momentum flux. There are three principle
choices: a surface conforming to the surface of the trapped particle, a surface of simple
geometry close to the particle enclosing it, and a spherical surface in the far field,
as shown in figure~\ref{integration_surface}.
The latter two of these are often the best choices. If a surface in the far field is
chosen, it can be possible to make far-field approximations to
simplify calculation of the momentum flux. If a nearby surface of simple geometry is chosen,
the same surface can be used for particles of different shapes, simplifying implementation.

If there are multiple particles within the trap, we will usually need to calculate
the optical force acting on each particle. This is important, for example, when
considering optical binding \citep{chaumet2001,chvatal}. In this case, we cannot find the
individual forces by integration of the total field in the far field. We can instead use surface
surrounding each individual particle, integration over each of which will yield the force acting
on the enclosed particle. In a multiple-particle T-matrix formulation, it is still possible to
use the usual single-particle summation formulae \citep{nieminen2014} to find the force, if we use
incident and scattered field mode amplitudes for each particle individually.

While there are many possible methods, they largely fall into three
groups: finite element methods (FEM), finite difference methods,
of which the most notable variants are the finite-difference time-domain
method (FDTD) and the finite-difference frequency-domain method (FDFD),
and approximate methods, notably Rayleigh scattering and geometric optics
or ray optics.

In the finite element method (FEM), the computational space is subdivided into
finite elements (Volakis et al., 1998). The values
relevant to the PDE inside or at the surface of each element are approximated by some known
function, perhaps linear or a higher order function \citep{volakis_fem}. The interaction between
each element or an element and its surrounding elements is described by a matrix that depends
on the particular definition and the chosen division of the computational space.
For scattering problems, FEM might be used to refer to a number of methods that involve solving
either sparse or dense systems of equations. The most important FEM as far as optical tweezers
is concerned is the the discrete dipole approximation (DDA). The physical interpretation
of DDA involves representing a large scattering particle by multiple smaller interacting
dipoles whose polarisability is known \citep{mishchenko2014book}.
The interaction between each dipole is described by a dense matrix; the resulting linear system
approximates the scattering by the combined object. \cite{yurkin2007b} provide a good
overview of DDA including recent developments and comparisons to other methods. Unlike
other FEM, DDA doesn't require the space surrounding the scatter to be discretised, unless the
surrounding space is inhomogeneous or contains other objects. DDA is more suitable for smaller
isolated particles and particles with smaller (relative) refractive indexes due to the requirement
to solve a dense linear system.

The finite-difference time-domain (FDTD) method refers to a method described by \cite{yee1966} for
solving systems of coupled partial differential equations. Although it was originally formulated for
solving the Maxwell equations, FDTD can also be applied to other systems of differential equations.
The original formulation of FDTD for the Maxwell equations involved calculating the electric
and magnetic fields at locations on a structured grid
spanning the computational space. Spatial derivatives in the Maxwell equations are calculated using
second order finite difference approximations involving the adjacent locations on the structured grid. The
fields are advanced through time using a leapfrog scheme with second order accuracy, where the
electric and magnetic fields are updated at alternate half integer time steps. Since Yee's original
method, there have been numerous improvements and specialisations such as unconditionally
stable methods or single step methods \citep{inan,raedt}.
One disadvantage of the FDTD method is difficulty in representing objects with smooth surfaces
using a structured Cartesian grid; when the surface doesn't conform to the Cartesian grid, this
introduces staircasing error \citep{inan}. The simplest approach is to increase the grid
resolution. However, this
results in a major increase in the computational requirements. Other alternatives include using
non-Cartesian grids such as spherical or circular grids, sub-gridding
certain regions or incorporating a local distortion near curved object
boundaries \citep{hastings}. Use of non-Cartesian grids requires calculation of the Jacobian
and correcting the FDTD update equations appropriately, special attention should also be given
to discontinuities in the mesh.

The finite-difference frequency-domain (FDFD) method is very similar to FDTD except it assumes
time harmonic solutions for the incident and scattered fields \citep{loke2007a}. FDFD is very
similar to FEM where only interactions between adjacent elements are considered.
This results in a linear system describing the scatter. Unlike FDTD, FDFD
performs the scattering calculation for only a particular frequency, while FDTD is able to calculate
the scattering of multiple frequencies simultaneously.

\begin{figure}[!htbp]
\centerline{\includegraphics[width=3in]{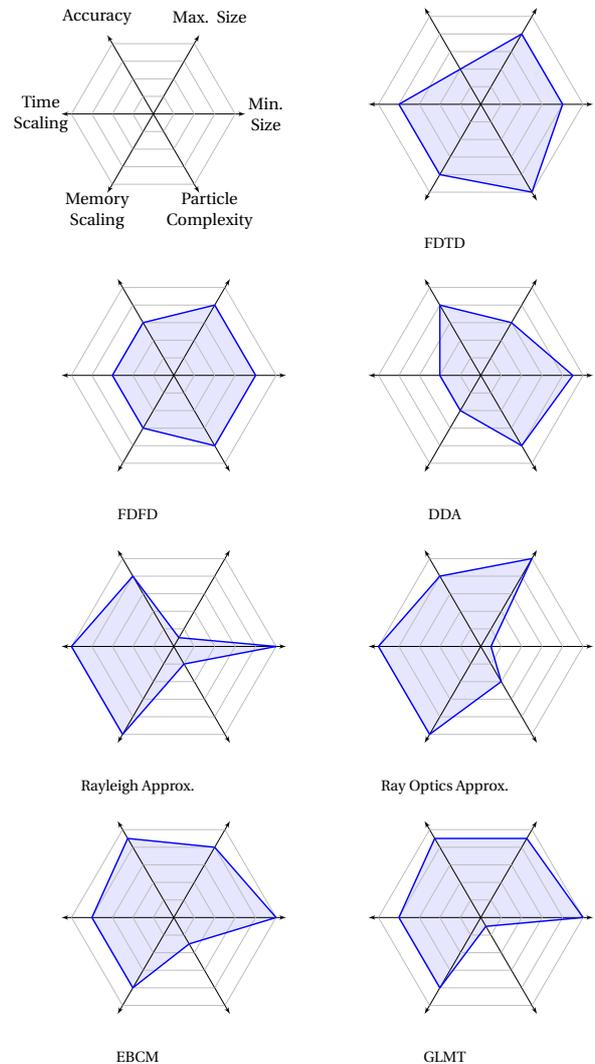}}
\caption{Comparison of different computational methods for simulating optical tweezers.
Further from the centre is better; e.g., the Rayleigh approximation, GLMT and EBCM are
the best of these methods for calculation forces on very small particles.
The particle complexity includes both the particle geometry and composition (inhomogeneity,
anisotropy, non-linearity). The accuracy
and computational requirements depend on the type of problem being solved, so this
comparison should be treated as a qualitative guide, rather than an exact quantitative
comparison.}
\label{isaacfig}
\end{figure}

The choice to use a particular computational method to model optical tweezers greatly depends
on the regime the problem falls into. For very large and very small particles the ray optics and
Rayleigh approximations are able to model particles with reasonable accuracy. DDA approaches
the Rayleigh limit for small particles but is able to simulate larger particles with fairly high
accuracy but scales relatively poorly with memory and time. The FDTD and FDFD methods
are able to simulate an extended range of particles but rely on being able to discretise the
computational space to describe fine details of the scatterer or rapidly changing fields. FDFD assumes
a particular form for the wave solutions, but is only able to simulate a single frequency;
in comparison, FDTD can easily deal with illumination such as short pulses, and can include
non-linear effects including frequency doubling, frequency mixing, etc.
A summary of these comparisons is presented graphically in figure~\ref{isaacfig}. The performance
and capabilities of the different methods depend on the type of problem being solved, so this
comparison should be treated as a qualitative guide, rather than an exact quantitative
comparison. This comparison is based on our own experience with these methods.

\section{Viscous drag}

Viscous drag is a key factor in the dynamics of a particle in an optical trap.
It determines the speed at which the optical forces and torques will move
or rotate the particle, and it also affects Brownian motion. For simulation
of optical tweezers, we wish to obtain the translational viscous drag tensor
for the particle (and the rotational drag tensor, if we need to include
rotation in the simulation). For the case of a spherical
particle, this is straightforward, since there is a simple analytical
solution: Stokes drag on a sphere. This gives
\begin{equation}
\mathbf{D} = 6\pi\eta a \mathbf{I},
\end{equation}
where $\mathbf{I}$ is the identity tensor. If the rotational drag tensor is
required, this is equal to
\begin{equation}
\mathbf{D}_\mathrm{rot} = 8\pi\eta a^3 \mathbf{I}.
\end{equation}

For cylinders, convenient formulae are given by \cite{gdlt}. For more general
cases, it can be necessary to resort to solving the fluid flow and calculating
the drag tensors. At low Reynolds numbers, the fluid flow is described by the
Stokes equation:
\begin{equation}
\eta \nabla^2 \mathbf{v} = \nabla p,
\end{equation}
where $\mathbf{v}$ is the velocity field of the fluid flow, and $p$ is the pressure field.
The two most promising approaches appear to be using the general solution in
spherical coordinates \citep{lamb1924b,pak2014} and direct finite-difference solution.

\subsection{Wall effects}

It should also be noted that nearby surfaces affect the viscous drag on a particle. In general,
this is a difficult problem \citep{happel}. However, the simple case of a sphere near a plane wall has a known
solution. The approximate solution by \cite{faxen} is often used. However, it is
only accurate when the particle is a large distance away from the wall, and fails when the
distance between the wall and the closest part
of the particle is less than 1 particle radius. That is, it cannot be used when it is most needed.
The exact solution presents serious difficulties in calculation. Fortunately, a simple and very
accurate approximation formula is available \citep{chaoui2003}.


\section{A recipe for simulation, part 2: Brownian motion}

Brownian motion in a viscous fluid, in the absence of other forces, can be
easily modelled. The probability distributions for displacements
of a spherical particle in each of
the $x$, $y$, and $z$ directions over a time interval $\Delta t$
are normal (Gaussian), with variance equal to $2D\Delta t$, where
\begin{equation}
D = \frac{k_B T}{6\pi\eta a}
\end{equation}
is the diffusion coefficient, and $k_B$ is Boltzmann's constant, $T$ is the
absolute temperature, $\eta$ is the (dynamic) viscosity, and $a$ is the
radius of particle. Thus, it is straightforward to simulate Brownian
motion using a Monte Carlo method. To calculate a displacement over $\Delta t$,
we can generate 3 normally distributed random numbers, $R_x$, $R_y$, and
$R_z$ with variance equal to 1, giving us
\begin{equation}
\Delta x = (2D\Delta t)^{1/2} R_x
\end{equation}
for the displacement in the $x$-direction, and similar results in the
$y$ and $z$ directions. This can then be repeated for subsequent
time steps, with new random numbers $R_x$, $R_y$, and
$R_z$ generated for each time step.

Notably, classical Brownian motion of this type is self-similar across
all time scales, i.e., fractal \citep{einsteinbook,nelson}, and consequently, the accuracy of
a Monte Carlo simulation like this is independent of the choice of
step size $\Delta t$. Therefore, if we aim to simulate a series of
measurements of particle position, it is sufficient to calculate
the particle position at only the times at which the position
is measured, and $\Delta t$ is the time interval between the
measurements. There is no need to calculate the position
for intermediate times.

However, in the presence of other forces, this changes. It becomes
necessary for the distance the particle moves in a single
time step to be small enough so that the other forces
do not change too much. Since the optical force
can change greatly over half a wavelength, the distance must
be a small fraction of the wavelength. Noting that a particle
of radius $1\,\mu$m will move, on average, a distance of
$1\,\mu$m in 2.1\,s due to Brownian motion in water at 300\,K,
and the distance scales with the square root of $\Delta t$,
we would need a time step of approximately $10^{-4}$s if
we want the distance moved to be less than 1\% of the wavelength.

\begin{figure}[!htbp]
\centerline{\includegraphics[width=3in]{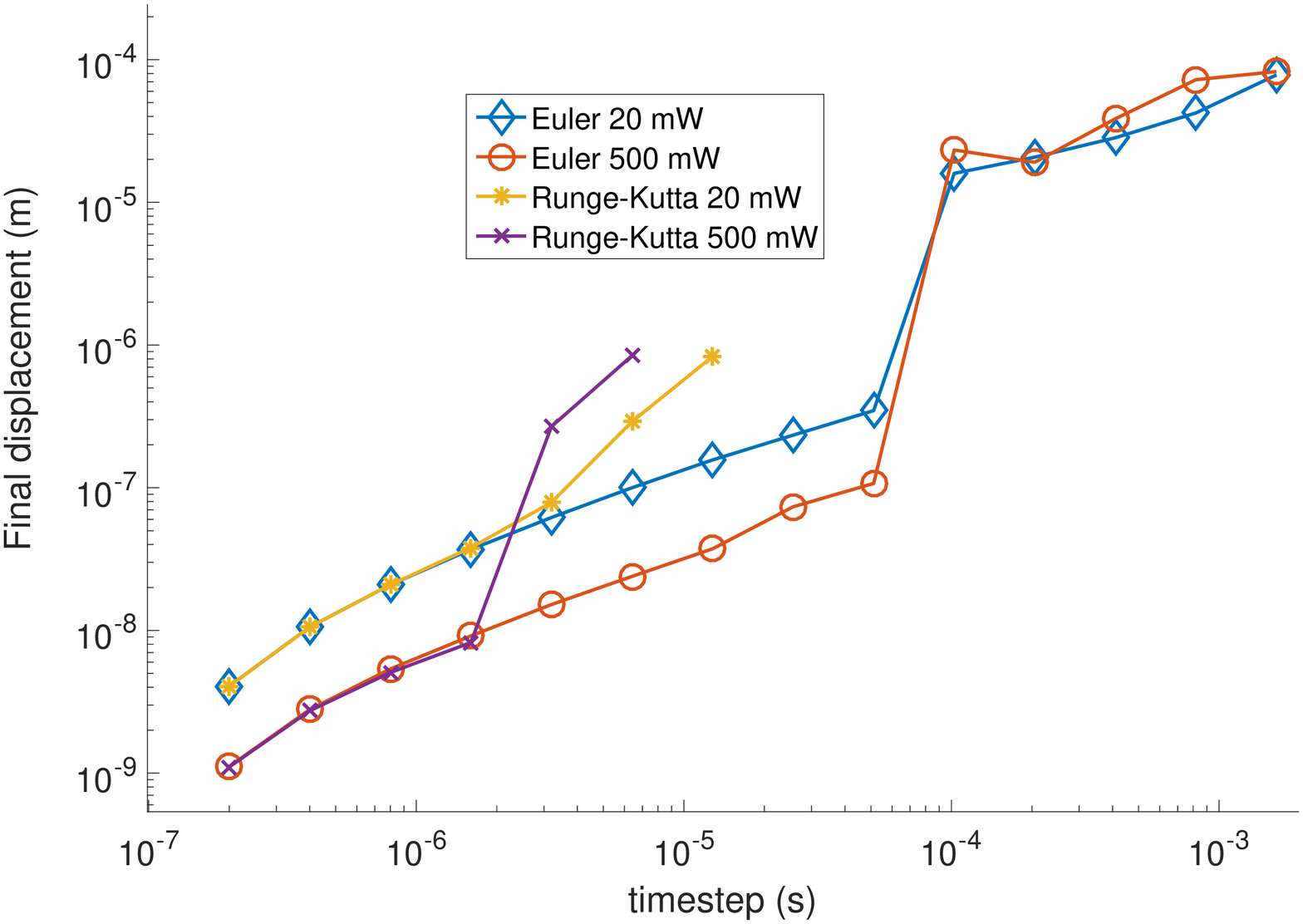}}
\caption{Convergence of final position of particle undergoing
Brownian motion in an optical trap as a function of
time step.}
\label{stepsizefig}
\end{figure}

We can investigate the effect of our choice of time step
quantitatively. We can generate a series of discrete Brownian
steps for a a time step $\Delta t_0$, and calculate the motion
of the particle. Then, we can double the time step, and sum
successive pairs of Brownian steps, to obtain half the number of
steps, each twice as long in time. This can be repeated, allowing
us to investigate the convergence of the calculation with
decreasing time step. The displacement over the time step due to
Brownian motion can be used to find an average velocity due to
Brownian motion over the time step; this can be expressed as
an average force over the time step and included in a predictor--corrector
method such as Runge--Kutta. An example is shown if figure~\ref{stepsizefig}
for a particle of radius $1\,\mu$m, comparing the convergence of
trajectories as the time step is reduced. The comparison includes
both Euler's method and a fixed-step 4th-order Runga--Kutta method.
The results indicate that
a time step of $10^{-4}$s (for a distance less than 1\% of the wavelength)
gives a reasonably small error. If higher accuracy is desired, a shorter
time step can be chosen. Due to the square root dependence of the 
distance, the scaling is relatively poor.

We recommend that a similar analysis of convergence be performed, especially if
simulations are sufficiently lengthy so that a just-small-enough
for acceptable error time-step should be chosen.


For very short time steps, the Stokes drag formula can be inappropriate
\citep{franosch,kheifets}, and for very short time steps, the transition
to ballistic motion becomes apparent \citep{huang_brownian}.
As long as the time step is not too short, this will not
present any difficulty. If simulations of a particle trapped in gas are being
performed, then the transition between ballistic and continuous regimes is
important \citep{li_brownian} at larger time steps.

\subsection{Nonspherical particles}

If the particle is nonspherical, the translation and rotational drag
tensors will be different in different directions, and the variance
of the Brownian motion will be different in different directions.

It is simplest to calculate the random Brownian steps in the rest frame
of the particle, and then transform the motion to the stationary frame.

The question of suitable time steps was left open earlier. From the
discussion of Brownian motion above, we see that Brownian motion
can be the main factor limiting our choice of time step.
However, it is wise to check that the optical force will not move the particle
too far during the time step. A maximum time step based on the optical
force can be found, and another based on the Brownian motion. The smaller
of the two can then be used as the actual time step for the computation.
If Brownian motion can be neglected, then it can be convenient to 
use an adaptive step size solver for initial value problems.
If we wish to calculate simulated measured at specific times,
we can choose time steps so that these specific times match
times at which we calculate the position of the particle in the trap.
Alternatively, it may be possible to interpolate between the calculations.

\section{Open questions}

A number of open questions and unsolved problems in simulation
of optical tweezers remain. We present a selection of them here,
in the spirit of presenting useful and interesting challenges to
those who wish to tackle them. Interesting work has been performed
on some of these topics, giving a hint of many interesting results
yet to be uncovered.

\begin{itemize}

\item \textbf{Optical force on complex particles.} While
optical forces and torques exerted on a wide range of particles
can be readily calculated, those on large and complex particles
remain challenging.

\item \textbf{Nonlinear particles.} Particles with non-linear
electromagnetic properties have the potential for many interesting
behaviours in optical traps \citep{pobre1997,pobre2006,devi}.

\item \textbf{Deformable particles.} These present a double challenge.
First, it is necessary to calculate not just the optical and viscous
forces acting on the particle, but also the stresses and consequent deformation
of the particle. Second, the deformation results in change in the optical
force. The deformable particles of most interest are red blood cells
\citep{li,dao,rancourtgrenier} and
other cells \citep{guck_biophysJ_2005}, but simpler objects such as vesicles, which
are sometimes used as simple analogs of cells, are also of interest
\citep{noguchi}.

\item \textbf{Wall effects on viscous drag on non-spherical particles.}

The movement of non-spherical particles near surfaces is important in many
biological systems. One interesting example is the motion of sperm,
which dramatically change in their swimming behaviour near surfaces.
\citep{elgeti,nosrati}. Optical tweezers offers an opportunity to
explore this behaviour, either by trapping sperm and measuring
swimming forces (as done for free-swimming sperm by \cite{nascimento})
or by trapping and moving analogs near surfaces. Simulations
would be very helpful for identifying changes in motion that result
from changed behaviour of sperm near surfaces; such simulations would
need to account for wall effects on the motion.

\item \textbf{Interaction between trapped particles and complex biological
environments.}

The work on deformation of red blood cells noted above can be considered a special
case of this. More generally, a trapped particle can interact with membranes,
macromolecules, cells, complex fluids, etc. Modelling its interaction with
such an environment can be challenging. Where
the behaviour of living cells needs to be included (e.g., swimming behaviour
of bacteria or sperm, ingestion of the trapped particle by a macrophage, etc.),
realistic models of the behaviour are required. This is a very complex
problem that remains largely untouched.

\item \textbf{Heating and thermal effects, including convective flow.}

While heating is often ignored in optical trapping simulations---wavelengths
and beam powers are often chosen to be such that absorption and consequent
heating in minimal, to avoid damage to live biological specimens---there
can be significant heating when absorbing particles are trapped.
Heating introduces a wide range of effects, from changes in the viscosity of
the surrounding fluid due to increased temperature, convection currents,
and effects such as thermophoresis \citep{floresflores}. If there
are liquid--liquid or liquid--gas interfaces present, the dependence
of surface tension on temperature can produce strong flows due to
Marangoni convection \citep{miniewicz}. In general, the temperature distribution
drives the convective flows, and the convective flows can alter the temperature
distribution, and also the position of the particle within the trap (thus altering
the absorption of light and the temperature distribution). This coupling
makes the solution of the problem difficult; an iterative method might be required.
The time scales involved can be investigated---if conduction dominates energy transport,
then it may be possible to ignore the effect of convection on the temperature distribution,
and the problem, while still challenging, is greatly simplified.

\end{itemize}


\section*{Acknowledgments}

This research was supported under Australian Research Council's
Discovery Projects funding scheme (project number DP140100753).




\end{document}